# Mechanism of formation and peculiarities of structure of bulk MgB$_2$ compound specimens

E.I. Kuznetsova[1], S.V. Sudareva[1,*], T.P. Krinitsina[1], Yu.V. Blinova[1], E.P. Romanov[1], Yu.N. Akshentsev[1], M.V. Degtyarev[1], M.A. Tihonovskiy[2], I.F. Kislyak[2]

[1]*Institute of Metal Physics UB RAS, S. Kovalevskaya Street 18, Ekaterinburg 620990, Russia*
[2] *National Scientific Centre Kharkov Institute of Physics and Technology, Academicheskaya 1, Kharkov 61108, Ukraine*

[*]sudareva@imp.uran.ru, +7-3433783679

**Abstract**

The existence of two phases within one and the same hexagonal lattice of MgB$_2$ compound, differing in Mg and B (in the homogeneity region) and especially in impurity oxygen content, as well as in microstructure, is demonstrated by various techniques. The regions corresponding to these two phases of MgB$_2$ have the sizes of 100-500 μm, and they fill the whole bulk of specimens, alternating with each other. It is suggested that the two-phase state of MgB$_2$ compound is caused by specific features of its formation mechanism (as a result of synthesis at 800-1000°C), including the stages of Mg melting, dissolution of solid boron in it up to the composition of MgB$_2$ and further crystallization of the MgB$_2$ compound from the melt with the formation of dendrite-like structure with corresponding redistribution of main components and impurities.

Keywords: superconductors, electron microscopy, microstructure

## 1. Introduction

Nowadays the MgB$_2$ compound attracts great attention of the researchers due to the possibility of its practical application as a superconducting material. This compound has the superconducting transition temperature of about 39 K which is out of the limit of BCS theory; it has a very high current carrier density of $(1.7–2.8) \times 10^{23}$ holes/cm$^3$, that is 2 orders of magnitude higher than that for YBCO and Nb$_3$Sn compounds; high critical current density, J$_c$ (4.2 K, 0 T) $> 10^7$ A/cm$^2$; high Debye temperature, $\Theta_D \sim 900$ K, and a very low resistivity in the normal state [1]. These characteristics indicate that MgB$_2$ has the potentiality of superconductive applications in high-power field and electronic devices, and it will be the best material to replace the traditional Nb alloy superconductors working at the liquid helium temperature. Besides, this compound is of interest for the science of superconductors.

A concept that pure single crystal $MgB_2$ combines properties of type 1 and 2 superconductors with two types of electron bonding and two superconducting gaps ($\Delta_1 = 2.2$ meV and $\Delta_2 = 7.1$ meV), referred to as a 'novel type – 1.5 Superconductivity', is analyzed in [2], taking into consideration the available literature data. Based on the two-component character of the $MgB_2$ compound, Moshchalkov et al. [2] calculated distributions of Abrikosov's magnetic vortexes in an external magnetic field of H = 5 Oe (where they look like unconventional stripes) and without magnetic field (the so-called arachnoid patterns). It is interesting to note that these vortex distributions to a great extent coincide with experimentally observed vortex locations by Bitter method. Note that no pronounced variations of Mg and B concentrations along the direction perpendicular to the magnetic vortex stripes (the length of about 100 µm) were detected in [2].

Bulk samples of $MgB_2$ compound or ribbons based on it are usually fabricated from precursors with Mg and B powders with the content ratio of 1:2 or 2:2 (or from the finished $MgB_2$ compound) by pressing of tablets or rolling of $MgB_2$-filled metallic tubes and further annealing (with enveloping in Ta or Zr foils or without it) in Ar, vacuum or under the pressure of Mg vapors at 800°C (most frequently) or sometimes at 650-750°C or 950-1050°C [3-10]. In some papers, see, e.g. [11], bulk $MgB_2$ samples were obtained by hot-pressing. We are not aware of any publications on the formation mechanisms of $MgB_2$.

The structure of $MgB_2$-based specimens (bulk specimens, ribbons, wires) was studied in a number of publications [3-6,8,10-15]. In the X-ray patterns of specimens with the Mg:B ratio of 1:2 there are practically always the lines belonging to two phases, $MgB_2$ + MgO. The MgO oxide is present in specimens synthesized in a wide temperature range from 600 to 1000°C [10]. As demonstrated by TEM [4], the coarse faceted crystals of $MgB_2$ may be located in the fine-grained MgO phase serving as a matrix, or, vice versa, fine and coarse crystals of MgO, sometimes faceted, are present in $MgB_2$ crystals. Crystallographic connection is possible between the $MgB_2$ and MgO phases. In [12], particles of various phases (Mg-B-O the sizes of about 0.2 µm, $MgB_4...MgB_{12}$) were found by SEM in bulk synthesized $MgB_2$ specimens. Such wide spread in precipitates composition may be due to chemical non-uniformity of specimens. The authors of believe these particles to serve as pinning centers.

Of special interest is publication [15], in which small and large particles (the sizes of 5-100 nm) were observed by TEM in $MgB_2$ crystals (bulk specimens). According to the analysis carried out in this study small particles (the sizes of less than 30 nm) have composition Mg-B-O and are enriched in oxygen compared to the surrounding $MgB_2$ matrix, whereas coarse particles belong to MgO. According to HREM, small Mg-B-O particles have the same lattice and orientation as the $MgB_2$ matrix. Modulated structure is observed in small particles, representing periodical alternation of B and O composition. At certain conditions boron is displaced by oxygen, which results in the formation of coarse MgO particles. Based on these data it may be concluded that the presence of MgO in synthesized specimens of Mg:O = 1:2 to a certain degree results from the phase transformation of $MgB_2 \rightarrow Mg(B,O)_2 \rightarrow MgO$.

In the present study the results of observation of large-scale foliation of the $MgB_2$ compound proper after synthesis at 800-1000°C into two phases or fractions are demonstrated. These phases have one and the same hexagonal crystal lattice with similar parameters $a$ and $c$ (in the X-ray patterns only single narrow lines of $MgB_2$ are observed in the range of $2\theta = 20$–$90°$), but they differ in Mg and B content (within the homogeneity region) and especially in oxygen content; they are spatially separated into large areas the sizes of 100-500 μm and posses different microstructure (dense grained or porous grained structure).

## 2. Experimental procedure

Three types of bulk $MgB_2$ specimens have been studied. Specimen No. 1 was fabricated from powders of Mg 98% pure and amorphous boron 99.98% pure with the atomic ratio of 1:2. The mixture of these powders was pressed into tablets of diameter about 10 mm and 3 mm thick. The tablets were placed into a thin-walled refractory steel tube with vacuum plug, with a capillary outlet for vacuum degassing at 300°C. The compound was synthesized at 820- 850°C for 2 h in Ar at the pressure of P ≥ 100 MPa.

Specimen No. 2 was made of 99.98 pure Mg and metallic boron melted from amorphous boron 96.93% pure. The mixture of Mg:B = 1:2 ratio was pressed into tablets. The $MgB_2$ compound was synthesized at 1000°C for 2 h in Ar at the pressure of 1 MPa.

Specimen No. 3 is also the tablet obtained as in case of specimen No. 1, but with Mg:B ratio of 1.5:2. The tablet was enveloped in Zr or Ta foils and placed into an iron ampoule. The vacuum degassing at 300°C was followed by the annealing at 800–850°C for 2 h in the vacuum.

Specimen No. 1 had the highest density (d = 2.43–2.45 g/cm$^3$) and high microhardness, $H_\mu$ = 760–930 kg/mm$^2$ [16].

We have specially chosen for our studies specimens with various nominal compositions fabricated by different techniques (synthesis temperatures, atmosphere and pressure being varied) to reveal any common elements of structure. The common feature for all the specimens considered is the high synthesis temperature (800-1000°C).

The as-synthesized specimens were studied by the following methods: X-ray analysis in Cu $K_\alpha$ radiation (automated diffractometer STADI-P, with sensitivity of 1-3 vol. %); scanning electron microscopy (Quanta-200 with EDAX for microanalysis); transmission electron microscopy (JEM-200CX) and measurements of temperature dependences of magnetic susceptibility (SQUID-magnetometer MPMS-XL-5XL, the measurements were carried out in alternating magnetic field of frequency 80 Hz and amplitude 4 Oe). Thin foils for TEM were made by electropolishing.

## 3. Results and discussion

### 3.1. X-ray studies

X-ray diffraction patterns of the specimens under study are shown in Fig. 1. The diffraction pattern of specimen No. 1 (with Mg:B ratio of 1:2) demonstrates two sets of lines belonging to $MgB_2$ and MgO phases. The lattice parameters of $MgB_2$ are $a = 0.30853$ nm and $c = 0.35250$ nm. X-ray density of $MgB_2$ is d = 2.6298 g/cm$^3$.

Specimen No. 2 (with Mg:B ratio 1:2) also contains two phases, $MgB_2$ and metallic Mg.

Specimen No. 3 (with Mg:B = 1.5:2) in spite of Mg excess possesses only two phases, $MgB_2$ and MgO. Thus, in the X-ray patterns of all synthesized Mg-B specimens the second phase, MgO(Mg) is always observed.

As a rule, X-ray patterns are given in literature for the angle range $2\theta$ from 20 to 90°, and relatively narrow lines of $MgB_2$ are observed. We have carried out the X-ray studies in the range of $2\theta = 20$–160°. It has been found that the (302) line of $MgB_2$ in the range of $2\theta = 150$–155° for all three specimens studied is markedly broadened, and the doublet of $K\alpha_1$ and $K\alpha_2$ lines is not resolved. This may be due to high internal stresses in a specimen, fine-dispersed structure or the presence of phases of different atomic composition within one crystal lattice of $MgB_2$. As demonstrated below, the latter is the determining factor.

### 3.2. Structure: scanning electron microscopy with microanalysis and transmission electron microscopy

Fig. 2a demonstrates SEM image of the structure of Specimen No. 1 (Mg:B = 1:2) taken in secondary electrons. We are not aware of observation of such structure in any other studies. On the dark background one can see coarse needles mostly propagating from one center. It is clearly seen at high magnification that these needles and areas between the needles consist of separate crystallites (Fig. 2b). As seen from the image of the area (Fig. 2a) taken in Mg $K\alpha$ radiation, the Mg content in the needles is considerably higher than in the regions between the needles (Fig. 2c). Different content of Mg in these two types of structure could be due to their different depth in the cross-section. However, examination of this area in back-scattered electrons has shown that the difference in depth is negligible.

Fig. 2d demonstrates variation of basic components Mg and B and impurities O and C along the scanning line coinciding with the base of B distribution line. It is seen that Mg concentration varies regularly, its raise coinciding with the location of needles and its drop with the regions between the needles. Relative to Mg, concentration of oxygen changes oppositely, namely, in the needles with high concentration of Mg concentration of O is lower whereas between the needles it is higher.

The above-described structure of $MgB_2$ resembles dendrite structure forming at crystallization from a melt. As a rule, initial dendrites forming at crystallization have an optimal chemical composition whereas excess and impurity atoms transfer to interdendritic areas. That is why secondary crystals forming in the interdendritic space possess another chemical composition. We consider the structure shown in Fig. 2 as dendrite-like, and we believe it to reflect a certain mechanism of $MgB_2$ formation from the melt. The data of spot microanalysis in many areas of the cross-sections of specimen No. 1 (up to 10-27 points) (Table 1) completely coincide with the above mentioned description, namely, there is more Mg and B in the needles and several times less oxygen than in the space between the needles.

A question may arise whether the higher magnesium content in the needles is associated with the presence of high amount of MgO in them, which is revealed in the X-ray pattern (Fig. 1). However, it is not so. As seen from Fig. 2c, distribution of Mg throughout the needles is practically homogenous, which means that the MgO particles are small. Besides, the higher content of Mg (40-51 at. %) in the needles is accompanied with the presence of comparatively low amount of oxygen in them (6-9 at. %). This means that the number of MgO particles in the needles is not larger than in the space between the needles. In this case it can be concluded that the enhanced amount of Mg in the needles is concentrated mainly in the $MgB_2$ lattice (its composition being closer to stoichiometry), whereas the $MgB_2$ phase in the space between the needles is depleted in magnesium.

The conclusion on the formation of two phases within one and the same crystal lattice $MgB_2$ is well founded. According to the review [17], the presence of cation non-stoichiometry in borides of a number of $s$-, $p$- and $d$-metals is a well-established fact. It was found that cation vacancies in $MgB_2$ lattice can arise at certain synthesis conditions and in case oxygen doping. According to [17], the $MgB_2$ phase may contain up to 5–9 % of cation vacancies, but the corresponding changes in lattice parameters are insignificant. The superconducting transition temperature lowers by 1-3 K with increasing concentration of vacancies in Mg-sublattice of $Mg_{1-}$ $_x$B$_2$ compound within its homogeneity region, $0 \leq x \leq 0.1$. Our X-ray data are in full agreement with that of [17]. Both MgB$_2$ phases, differing in Mg and B content (in the amount of vacancies) and in impurity oxygen concentration have similar lattice parameters. In the range of $2\theta = 20–90°$ the narrow single lines of MgB$_2$ are observed in the X-ray patterns. A noticeable line broadening due to a small difference in lattice parameters of two MgB$_2$ phases is observed only in the range of $2\theta = 150–155°$, see Fig. 1. We believe that vacancy formation is favored by the impurity oxygen.

Images of the structure of specimen No. 2 (Mg:B = 1:2) taken in backscattered and secondary electrons (BSE and SEI regimes) markedly differ from those of specimen No. 1 in the shape of fragments and the scale (Fig. 3a,b). However, as demonstrated below, the two-phase states of MgB$_2$ in these two specimens are similar. Two types of areas are clearly seen in Fig. 3a: smooth dense areas without a visible contrast (shown by a circle) and loose areas consisting of separate crystals, many of which have striate structure (shown by a square). In the course of the cross sections preparation the loose phase crumbles out and shallow cavities are formed. Comparing Figs. 3a and 3b, one can see that smooth dense areas in Fig. 3a correspond to dark areas in Fig. 3b, and loose areas, which sometimes have a striated contrast, correspond to bright areas in Fig. 3b. Note that in [5,6] the similar contrast as in Fig. 3b was observed in SEM images of bulk MgB$_2$ specimens, where it was explained by the presence of open and closed holes. The pores are also clearly seen in Fig. 3b (they are shown with white arrows), but there are not many of them.

Let's analyze the structure of this specimen in more detail. As clearly seen in Fig. 4b, which demonstrates an area from Fig. 4a at higher magnification, the smooth dense areas (shown by a circle) also consist of separate crystallites, but the crystals in them are more closely connected forming dense conglomerates or sometimes coarse monoliths (see Fig. 4c). In Fig. 4b one can see loosely coupled crystallites of porous region (a brighter area on the right). The microanalysis data (Table 1) show that Mg and B content in the smooth dense areas are higher and O content is several times lower compared to the loose areas. It should be emphasized that this situation is realized in specimen No. 2, where, according to the X-ray data, the MgO phase is practically absent (Fig. 1). Hence, it may be concluded that the impurity oxygen is located in two MgB$_2$ phases in form of solid solutions. We attribute the smooth dense areas to the initial crystallization and the loose areas to the secondary crystallization (dendrite-like structure).

Of interest is the origin of bright striated loose areas of MgB$_2$ (Fig. 4d). Spot microanalysis inside the stripes and on their boundaries shows that there is practically no difference in the content of basic elements and impurities. Examination of this area in BSE regime reveals some difference between the stripes and their interfaces resulting from the surface relief: interfaces are located on the relief crests, and they seem to be harder than the internal regions. The structure shown in Fig. 4d resembles in appearance the structure of columnar $Nb_3Sn$ grains which forms as a result of diffusion interaction between Nb-filaments and Cu-Sn matrix in bronze-processed superconducting composites and is attributed to a directed diffusion flow of Sn [18].

Thus, in this case as well (specimen No. 2) at least two $MgB_2$ phases with different content of basic components (within the homogeneity region) and impurity oxygen (Table 1) are observed. The great amount of the impurity oxygen in all specimens is in agreement with the available experimental data of other authors (see, e.g., [8]). As mentioned in [8], it is an important but quite a difficult problem to find out at what stage and how this oxygen is introduced into the $MgB_2$ samples.

In spite of an enhanced Mg content in the precursor (Mg:B = 1.5:2), the structure of specimen No. 3 resembles that of specimen No. 2 (compare figs. 2 and 5), though Mg in the former is in an oxidized state, see Fig. 1. Two types of structures are also observed here, dark smooth dense areas with an enhanced content of Mg and B and markedly reduced content of oxygen (see Table 1 and Fig. 5a,b, (shown by a circle), and bright areas of irregular shape mostly with loose structure (Fig. 5a, shown by a square). In Fig. 5c, Mg distribution in dark smooth dense areas of specimen No. 3 is shown (the area shown in Fig. 5a). It is seen that the distribution is uniform, there are no coarse MgO particles, and the $MgB_2$ phase with an enhanced Mg and B content (close to stoichiometry) and small amount of oxygen is mainly present. But there are also regions with "intermediate" structure in the interdendritic areas of this specimen (Fig. 5d, shown by a triangle). These areas consist of clearly defined grains more closely adjoined to each other than in the loose areas, and Mg and O content in them is highly increased (about 50 at.% Mg and 40 at.% O). We believe these areas to be a mixture of $MgB_2$ and MgO phases, the latter constituting a considerable portion.

Thus, all the specimens under study fabricated from various precursors in different conditions demonstrate one common feature – they consist of two hexagonal $MgB_2$ phases with markedly differing oxygen content (to about 4 times). The observation of the two-phase state of $MgB_2$ is confirmed by the X-ray study: for all three specimens studied the (302) $MgB_2$ line in the range of $2\theta = 150–155°$ is noticeably broader than the instrumental widths of $K\alpha_1$ and $K\alpha_2$ lines (fig. 1b), which may be due to the presence of $MgB_2$ phases with somewhat different parameters. The treatment of the experimental X-ray patterns by the method of full-profile analysis with the PowderCell 2.4 computer program allows to estimate approximately the lattice parameters of these two phases as $a = 3.0795$Å and $c = 3.5194$Å (one phase), and $a = 3.0741$Å and $c =$ 3.5077Å (another phase). Large (up to 500 μm) dark dense areas and bright loose regions correspond to these two phases. Besides, the dense areas (formed as a result of initial crystallization) in specimens with component ratios of Mg:B = 1:2 and/or 1.5:2 have an enhanced amount of magnesium and boron compared to that in bright loose regions (resulting from the secondary crystallization). As reported in [8], 'the wires with pre-reacted $MgB_2$ (ex situ) show oxygen-poor $MgB_2$ colonies (a colony is a dense arrangement of several $MgB_2$ grains) embedded in porous oxygen-rich matrix introducing structural granularity', but the structure shown (SEM data) is not similar to that observed in the present study.

In TEM studies most often the areas corresponding to two types of $MgB_2$ structure are observed (Fig. 6, specimen No. 1). These are areas with coarse crystals and a fine-dispersed constituent (Fig. 6a) and regions with mainly fine-dispersed structure (Fig. 6b). In the electron diffraction patterns of the former (Fig. 6a, insert) one can see discontinuous Debye rings and sharp point reflections of $MgB_2$. Electron diffraction patterns of the fine-dispersed structure demonstrate strong continuous Debye rings and weaker point reflections of $MgB_2$ (Fig. 6b, insert). It is difficult to reveal reflections of MgO in the electron diffraction patterns, because the strong reflections of this phase with d = 0.210 nm and d = 0.1485 nm practically coincide with those of $MgB_2$ (d = 0.212 nm and d = 0.147 nm). It may be suggested that areas with coarse crystals are connected with the initial needles (Fig. 2a) and the fine-dispersed structure corresponds to the "interdendritic" space. The structures observed in the present study are basically the same as in [13,14].

All the specimens studied are superconductors with high critical temperatures $T_c$. Physical methods of investigation do not reveal the obvious presence of only these two phases of $MgB_2$ (there is some scattering in composition), see Fig. 7a-c. In the temperature dependences of magnetic susceptibility there are broadened superconducting transitions: for specimen No. 1 $\Delta T_c$ = 38–40 K (plus a small step at about 37 K), and for specimen No. 2 $\Delta T_c$=35–38 K. The temperature dependence of $\rho/\rho_{50}$ (for specimen No. 3) demonstrates a drastic drop of resistivity (at 37 K), which finishes with a small "tail" in the range of 37–38.2 K.

### 3.3. Plausible mechanisms of $MgB_2$ formation

If the common specific feature of the structure of specimens under study, namely, the presence of two phases $MgB_2$ differing in the content of basic components (within the homogeneity region) and especially impurities (oxygen) is considered, then it may be assumed that the synthesis process includes the stage of crystallization of $MgB_2$ from a melt which is completed with the dendrite-like structure with corresponding redistribution of main components and impurity oxygen. Consequently, the main mode of $MgB_2$ formation is the liquid mechanism.

As the melting temperature of Mg is 649°C and that of B is 2100-2200°C, at the synthesis temperature of 800-1000°C magnesium is melted (see the phase diagram in [19]). As seen from the phase diagram, magnesium may be in gaseous state at 1100°C and the pressure of 0.1 MPa. Therefore, at synthesis temperatures of 800-1000°C magnesium is both in liquid and gaseous states, as it has a high steam tension.

The liquid mechanism of $MgB_2$ formation may be conceived as follows. Solid boron dissolves in liquid magnesium till the melt concentration approaches the $MgB_2$ composition. Then crystallization starts with the formation of initial crystals with optimal chemical composition. All the "excess" atoms including gas impurities outgo to the space between the initial crystals, the "interdendritic" space. The secondary crystallization in this "interdendritic" space occurs at somewhat different composition of the melt, and the secondary $MgB_2$ crystals appear to be enriched in oxygen and depleted of the basic components (see Table 1). It is the cause of formation of two types of $MgB_2$ with slightly differing lattice parameters, but with different microstructure and chemical composition. A similar liquid mechanism was suggested for the Bi,Pd-2223 HTSC phase formation in [20]: the Bi,Pb-2223 is forming from a precursor by dissolution of the solid Bi,Pb-2212 and other oxides in the drops of eutectoid liquid of the 2223 composition (at T > 825°C). This mechanism was experimentally confirmed in [21], where the formation and spreading of the liquid phase and of the 2223 phase plates was observed in the scanning electron microscope equipped by a high-temperature device with heating from 25 до 840°C.

However, the initial crystals formation in the form of needles (specimen No. 1), propagating mostly from one center, requires, in our opinion, another explanation. A theory of solid-phase reaction was developed in [22-24]. It is based on experimental data concerning high mobility of metal ions in crystals with pronounced deviation from stoichiometry. Diffusion in such matters occurs by jumping on vacancy positions. The theory considered found its experimental confirmation, for instance, in [25], where the needles of $\alpha$-$Cu_2Se$ formed on a Cu substrate contacting with selenium ($T_m = 797$°C) at 450-600°C. This reaction is characterized by pronounced anisotropy of the rate of a compound crystal growth. The main diffusion of Cu atoms through vacancies in $\alpha$-$Cu_2Se$ occurs along the needles, the rate of the process along needle crystals being very high. This is due to the most favorable diffusion conditions, as the needle axes coincide with planes most densely populated by atoms of Cu and Se.

In our opinion, the foregoing considerations may be applied to the case of the occurrence of $MgB_2$ needles in specimen No. 1 at 800–850°C. The whole process of their formation may be conceived as follows. Magnesium vapor or its liquid fraction interacts with boron solid fraction. A layer of solid $MgB_2$ compound is formed on a solid boron surface serving as a sort of a substrate. This layer contains vacancies in Mg sub-lattice within the homogeneity region, and Mg ions start to diffuse through it by vacant positions at a high rate. Reaching an opposite side of the crystalline $MgB_2$ layer, Mg ions form needle crystals with boron in different directions inheriting the polycrystalline substrate crystallography. It may be assumed that in this case the development of such mechanism is affected by the presence of Ca impurity (Table 1).

Thus, based on the experimental structural data, two probable mechanisms of $MgB_2$ formation are suggested, which result in externally different but essentially similar structures, namely, two types of $MgB_2$ phases with slightly differing lattice parameters, but with different chemical composition of main components and oxygen, and different microstructure.

## 4. Conclusions

As a result of studies of structure of Mg:B = 1:2 and 1.5:2 specimens by various techniques it has been found that at synthesis (at 800–1000 °C) including magnesium melting, dissolution of solid boron in it and subsequent crystallization of $MgB_2$ from the melt actually two $MgB_2$ phases or fractions with different composition of basic elements (within the homogeneity region) and impurities (oxygen) are formed within one hexagonal crystal lattice. These two phases slightly differ in lattice parameters and markedly differ in their microstructure. One of them is denser and harder and contains optimal amount of Mg and B with small addition of impurity oxygen; it results from the initial crystallization. Another one, which forms at secondary crystallization, is looser, less hard and contains reduced amount of Mg and B and great amount of oxygen.

We believe that with the specific mechanism of $MgB_2$ formation including melting of Mg, dissolution of solid boron in it and subsequent crystallization of the $MgB_2$ compound the formation of two $MgB_2$ phases is quite possible even in the high-purity material. These phases differ in Mg content within the homogeneity region of $MgB_2$, they have different concentration of vacancies in Mg-sublattice and are spatially divided into a great number of comparatively small areas, and their superconducting temperatures somewhat differ as well.

### Acknowledgment

The work has been done on the equipment of the Collective Use Center of IMP, Ural Branch of RAS, within the RAS program, with partial support of the Presidium of RAS (project No. 12-P-2-1015).

## References

[1]    YiBing Zhang and Shiping Zhou (2010). $MgB_2$-MgO Compound Superconductor, Superconductor, Adir Moyses Luiz (Ed.), ISBN: 978-953-307-107-7, InTech, Available at: http://www.intechopen.com/articles/show/title/mgb2-mgo-compound-superconductor

[2]    Moshchalkov V, Menghini M, Nashio T, Chen QH, Silhanek AV, Dao VH,. Chibotaru LF, Zhigadlo ND, Karpinski J. Type-1.5 Superconductivity. Phys. Rev. Lett. 2009;102:117001.

[3]    Rajput S, Chudhary S. On the Superconductivity in in-situ synthesized $MgB_2$ tapes. J. Phys. Chem. Solid. 2008;69:1945-50.

[4]    Zhu Y, Wu L, Volkov V, Gu G, Moodenbaugh AR, Malac M, Suenaga M, Tranquda J. Microstructure and structural defects in $MgB_2$ superconductor. Physica C 2001;356:239-53.

[5]    Xu A, Ma Y, Zhang X, Li X, Nishijima G, Awaji S, Watanabe K. Superconducting properties of $MgB_2$ bulks processed in high magnetic fields. Physica C 2006;445-448: 811-3.

[6]    Liu CG, Yan G, Du SJ, Xi W, Feng Y, Zhang PX, Wu XZ, Zhou L. Effect of heat-treatment temperatures on density and porosity in $MgB_2$ superconductor. Physica C 2003;386:603-6.

[7]    Prikhna TA, Gawalek W, Savchuk YM, Sergienko NV, Moshchil VE, Wendt M, Zeisberger M, Habisreuther T, Dou SX, Dub SN, Melnikov VS, Schmidt Ch, Dellith J, Nagorny PA. Formation of magnesium diboride-based materials with high critical currents and mechanical characteristics by high-pressure synthesis. J. Phys.: Conf. Ser. 2006;43:496-9.

[8]    Birajdar B, Eibl O. Microstructure-critical current density model for $MgB_2$ wires and tapes. J. Appl. Phys. 2009;105:033903.

[9]    Alecu G, Cosac A, Zamfir S. Superconductivity in $MgB_2$. Annals of the University of Craiova, Electrical Engineering Series. 2008;30:382-5.

[10]   Chauhan SR, Chaudhary S. On the Residual Resistivity Ratio in $MgB_2$ Superconductors. IEEE Trans. Appl. Supercond. 2010;20:26-32.

[11]   Indrakanti SS, Nesterenko VF, Maple MB, Frederick NA, Yuhasz WM, Li S. Hot isostatic pressing of bulk magnesium diboride: mechanical and superconducting properties. Phil. Mag. Lett. 2001;81:849-57.

[12]   Prikhna T, Gawalek W, Eisterer M et. al. The effect of high-pressure synthesis on flux pinning in MgB2-based superconductors. Physica C 2012;479:111-4.

[13]   Li JQ, Li L, Zhou YQ, Ren ZA, Che GC, Zhao ZX. Structural properties of $MgB_2$

superconductors with a critical current density greater than $10^5 A/cm^2$. Chin. Phys. Lett. 2001;18:600-3.

[14] Asthana A, Matsumoto A, Kitaguchi H, Matsui Y, Hara T, Watanabe K, Yamada H, Uchiyama N, Kumakura H. Structural-microstructural characteristics and its correlations with the superconducting properties of in situ PIT-processed $MgB_2$ tapes with ethyltoluene and SiC powder added. Supercond. Sci. Technol. 2008;21:115013.

[15] Liao XZ, Serquis A, Zhu YT, Huang JY, Civale L, Peterson DE, Mueller FM, Xu HF. $Mg(B,O)_2$ precipitation in $MgB_2$. J. Appl. Phys. 2003;93:6208-15.

[16] Kislyak IF, Tihonovskiy MA, Malyhin DG et al. Studies of superconductivity of bulk $MgB_2$ and ex situ-wires Fe(steel)/$MgB_2$. Voprosy Atomnoy Hauki I Tehniki (Problems of Atomic Science and Technology) [in Russian] 2009;6:107-10.

[17] Ivanovskiy AL, Shein IR, Medvedeva N. Non-stoichiometric borides of s-, p- and d-metals: synthesis, properties and modeling. Russ. Chem. Rev. 2008;77:467-77.

[18] Deryagina IL, Popova EN, Romanov EP, Dergunava EA, Vorob'eva AE, Balaev SM. Evolution of the nanocrystalline structure of $Nb_3Sn$ superconducting layers upon two-stage annealing of Nb/Cu-Sn composites alloyed with titanium. Phys. Met. Metallogr. 2012;133:391-405.

[19] Liu ZK, Schlom DG, Li Q, Xi XX. Thermodynamics of the Mg-B system: implications for the deposition of $MgB_2$ thin films. Appl. Phys. Lett. 2001;78:3678-80.

[20] Morgan P.E.D., Housley R.M., Porter J.R. and Ratto J.J. Low level mobile liquid droplet mechanism allowing development of large platelets of high-$T_c$ "Bi-2223" phase within a ceramic. Physica C, 1991. V.176, No. 1-3. P.279-284.

[21] Maroni V.A., Teplitsky M., Rupich M.W. An environmental scanning electron microscope study of the Ag/Bi-2223 composite conductor from 25 to 840°C. Physica C, 1999. V.313, No. 3-4. P.169-174.

[22] Wagner C, Schottky W. Theoric der geordueten Mischphasen. Ztschr. Phys. Chem. B. 1930;11:163.

[23] Tretyakov YD. Solid-state reactions. Soros Educational Journal [in Russian]. 1999;4:35-9.

[24] Schamlzried H. Chemical Kinetics of Solids. Weinheim: VCH, 1995.

[25] Arharov VI, Esin VO. On the problem of mechanism of reaction diffusion in Cu-Se, Cu-Te, Ag-Se and Ag-Te. Fiz. Met. Metalloved. (Physics of Metals and Metallography) [in Russian] 1957;5:246-50.

Table 1. Results of microanalysis of MgB$_2$ specimens[*]

| Specimen number | Structure | Concentration, at.% | | | |
|---|---|---|---|---|---|
| | | B | Mg | O | Other impurities |
| No. 1 Mg:B = 1:2 | Initial crystals (bright needles) | 28–48 | 40–51 | 6–9 | C |
| | «Interdendritic» areas | 22–25 | 31–39 | 29–37 | C, Ca |
| No. 2 Mg:B = 1:2 | Initial crystallization areas (dark, smooth) | 36–49 | 43–46 | 7–16 | Si |
| | Secondary crystallization areas (bright, loose) | 24–26 | 20–31 | 39–53 | Si |
| No. 3 Mg:B = 1.5:2 | Initial areas (dark, smooth) | 34–45 | 37–47 | 9–14 | C |
| | Secondary areas (bright, loose) | 15–29 | 26–38 | 33–42 | C |

[*]Relative concentration limits of specimen components are given (from 10-27 measurements for every specimen); in every point the total concentration of all elements was taken as 100%. The atomic concentrations of boron are somewhat reduced, because the sensitivity of the method to this element is not high enough.

**Figure captions**

Fig. 1. X-ray diffraction patterns (taken in Cu Kα radiation) in the angle range of $2\theta = 20\text{-}160°$ for 3 specimens of $MgB_2$ fabricated from different precursors in different conditions. Numbers in the figure are the numbers of specimens.

Fig. 2. Scanning electron microscopy and microanalysis of $MgB_2$ (Mg:B = 1:2), specimen No. 1: a, b – secondary electron images (SEI); c – the area shown in fig. 2a in Mg radiation; d – distribution of basic elements (Mg and B) and impurities (O and C) along the scanning line (coinciding with the base of B distribution line).

Fig. 3. Scanning electron microscopy of $MgB_2$ (Mg:B = 1:2), specimen No. 2: a - backscattered electron image (BSE); b – SEI of the area shown in Fig. 3a.

Fig. 4. SEM image of specimen No. 2: a, b – SEI; c – SEI, structure of dark dense areas; d – structure of bright loose areas.

Fig. 5. Scanning electron microscopy and microanalysis of $MgB_2$ (Mg:B = 1.5:2), specimen No. 3: a – SEI; b – SEI, the structure of dark dense areas; c – SEI, the structure of dark and bright areas and areas with "intermediate" structure ($MgB_2 + MgO$); d – the area shown in fig. 5a in Mg radiation.

Fig. 6. Transmission electron microscopy of $MgB_2$, specimen No. 1: bright-field images and electron diffraction patterns (EDPs). Reflections belonging to $MgB_2$ are indicated by arrows in the EDPs.

Fig. 7. Temperature dependences of magnetic susceptibility of specimens No. 1 (a) and 2 (b) and the relative resistivity $\rho/\rho_{50}$ of specimen No. 3 (c).

# Figures

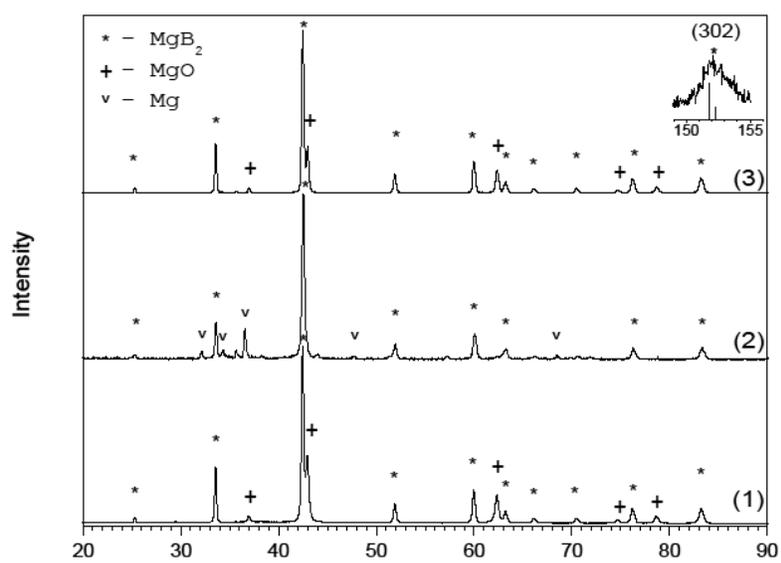

Fig. 1

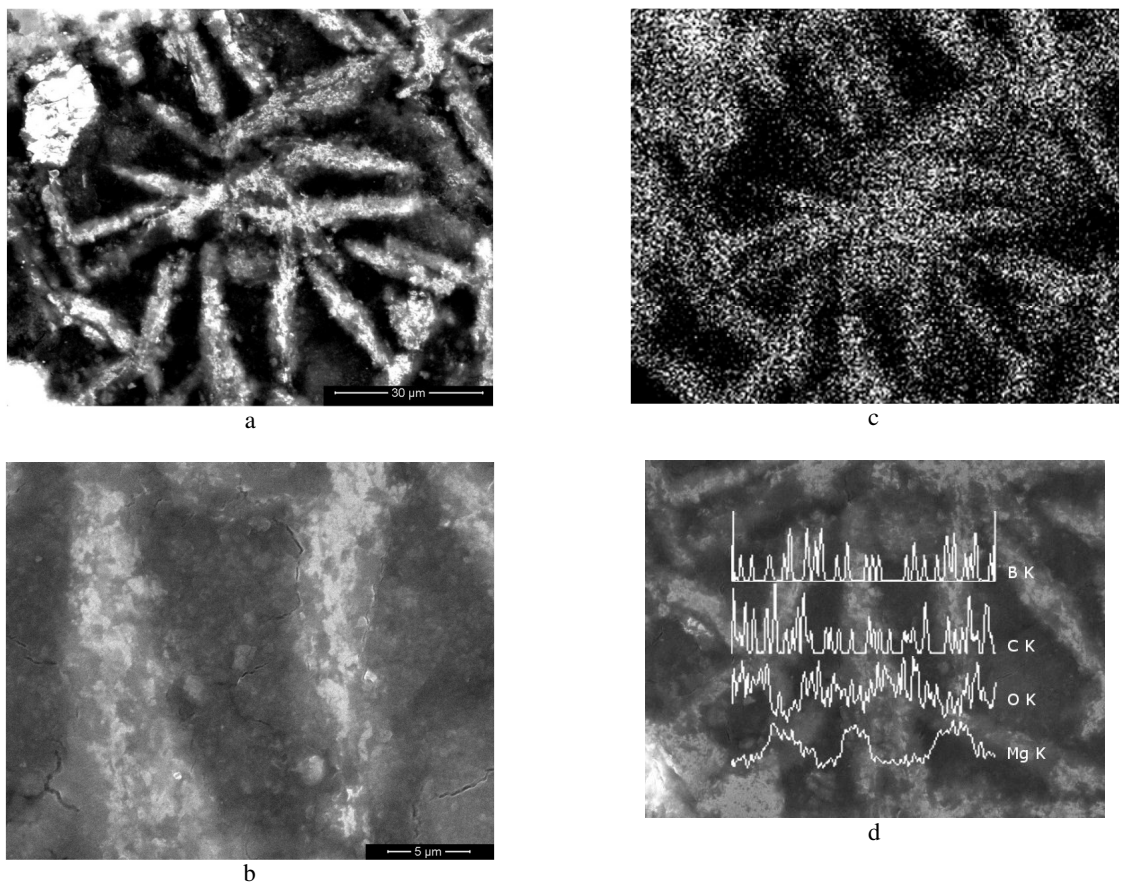

Fig. 2

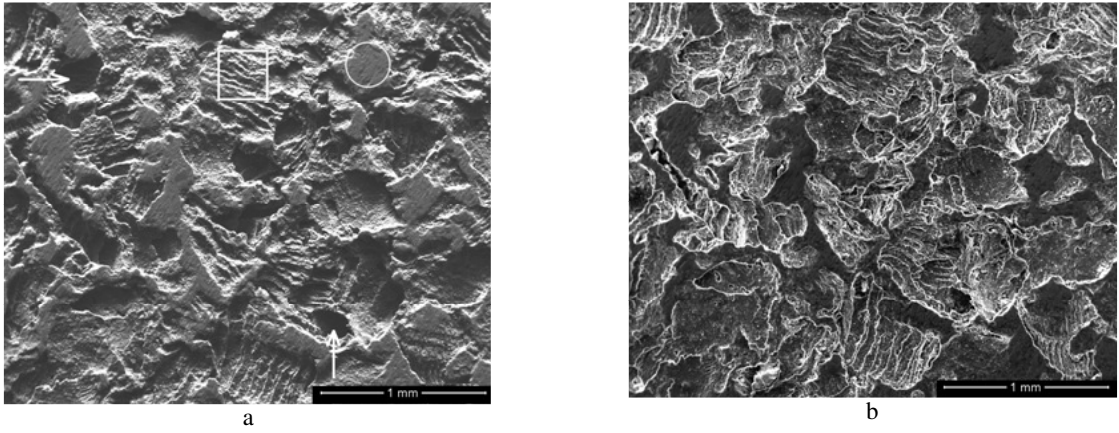

Fig. 3

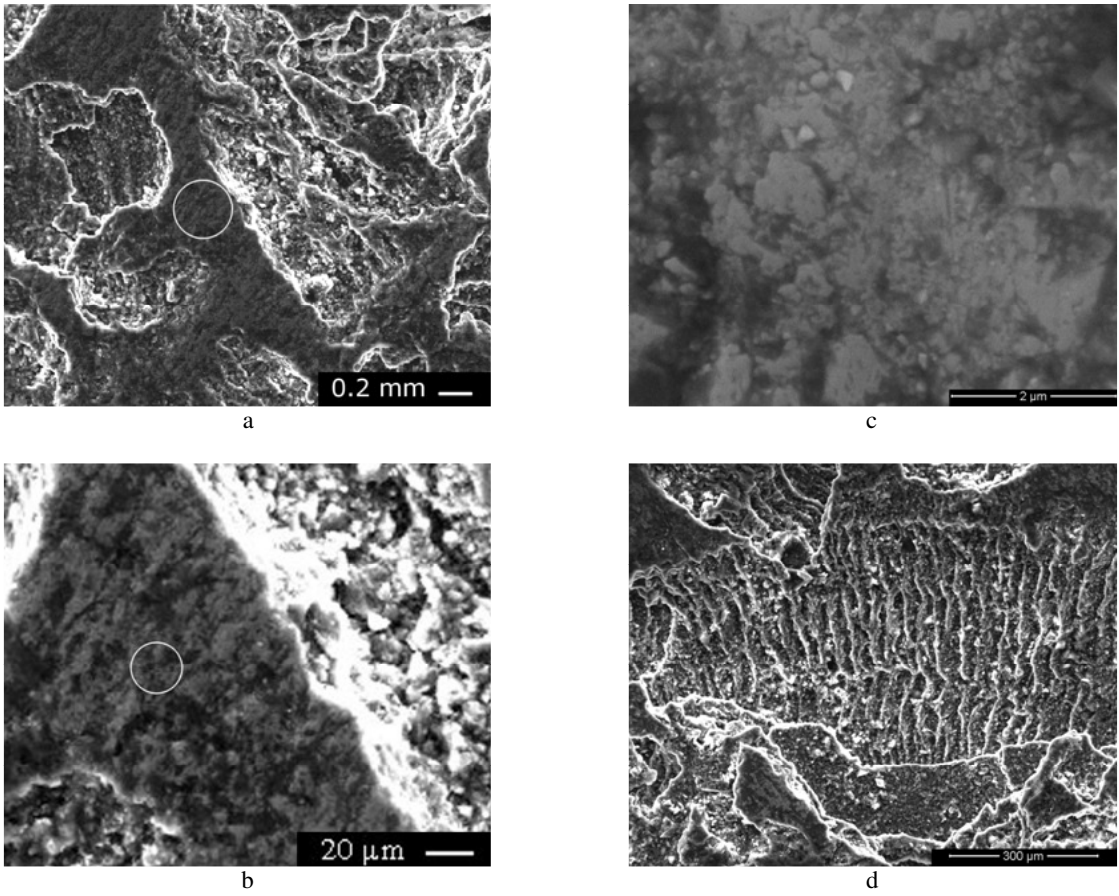

Fig. 4

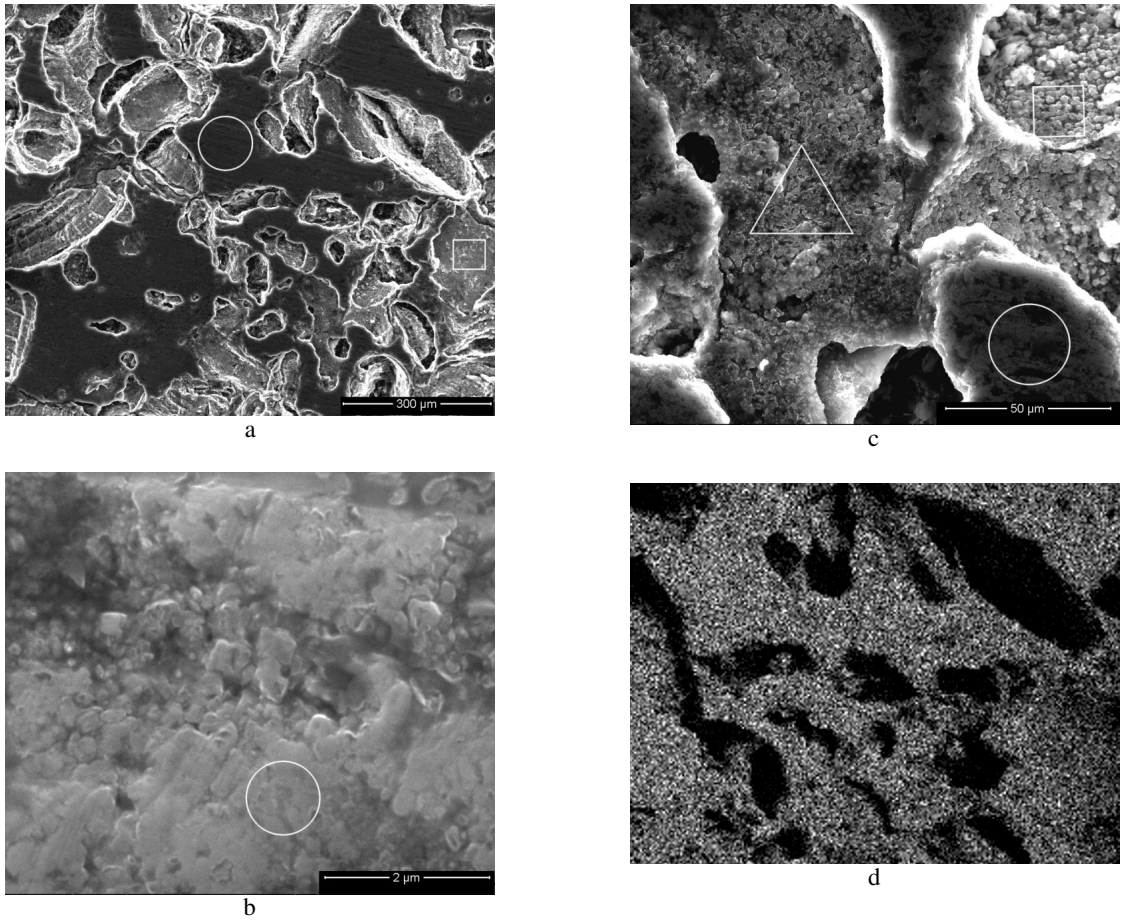

Fig. 5

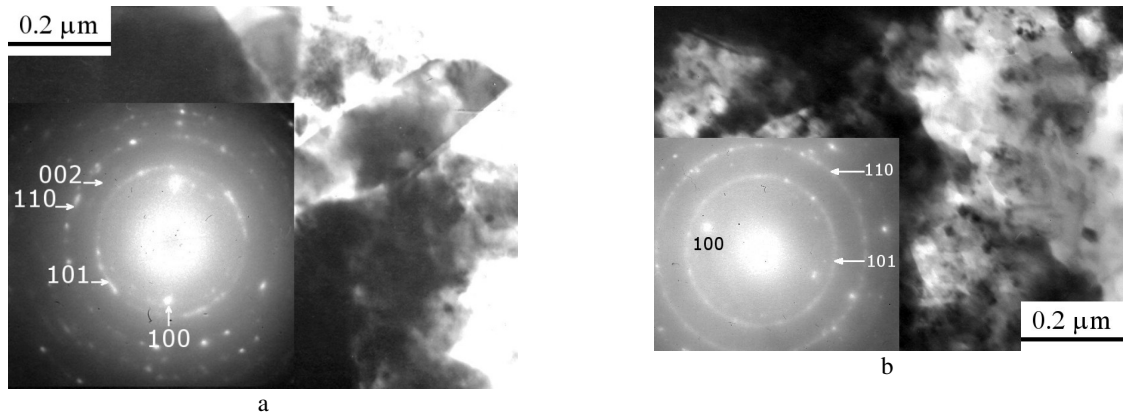

Fig. 6

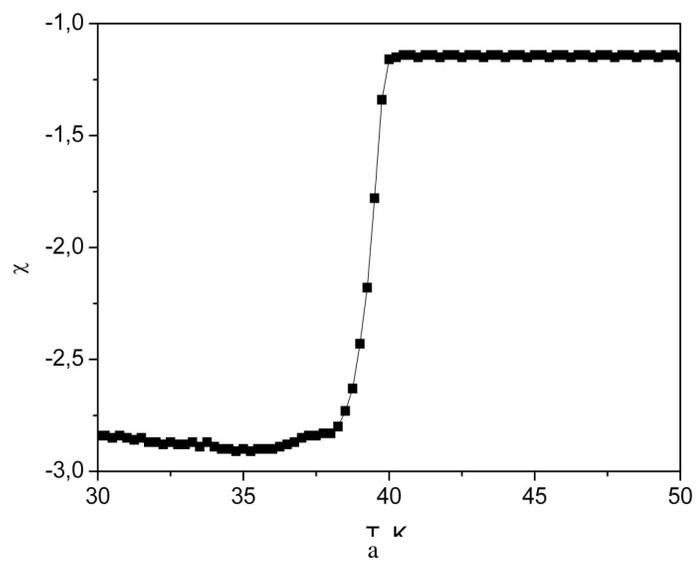

a

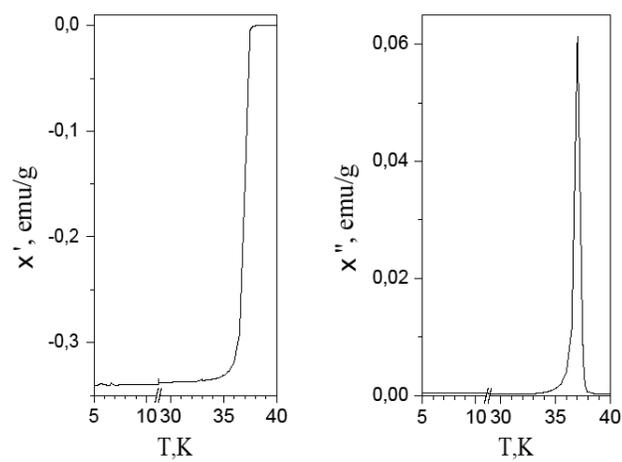

b

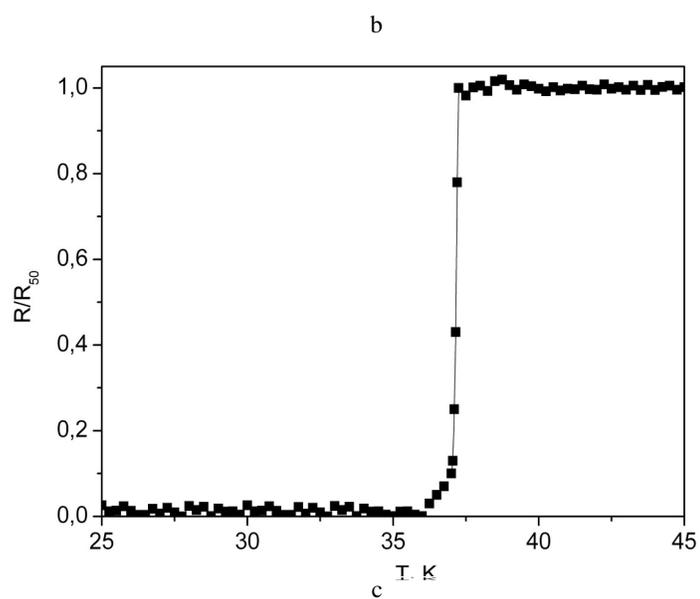

c

Fig. 7